\documentstyle[12pt,aaspp4]{article}
\newcommand{\NH}{N_{\rm H}}
\newcommand{\m}{\,{\rm m}^{-2}}
\newcommand{\FeKa}{Fe~K$\alpha$}
\newcommand{\Lsun}{\,L_\odot}

\lefthead{C. Simpson}
\righthead{Obscuration in NGC~3281}

\begin{document}

\title{Near-infrared and X-ray obscuration to the nucleus of the Seyfert~2
galaxy NGC~3281}

\author{Chris Simpson}
\affil{Subaru Telescope, National Astronomical Observatory of Japan,
650 N.~A`Oh\={o}k\={u} Place, Hilo, HI 96720}

\begin{abstract}
We present the results of a near-infrared and X-ray study of the Seyfert~2
galaxy NGC~3281. Emission from the Seyfert nucleus is detected in both
regions of the electromagnetic spectrum, allowing us to infer both the
equivalent line of sight hydrogen column density, $\NH =
71.0^{+11.3}_{-12.3} \times 10^{26}\m$ and the extinction due to dust,
$A_V = 22 \pm 11$\,magnitudes (90\% confidence intervals). We infer a
ratio of $\NH/A_V$ which is an order of magnitude larger than that
determined along lines of sight in the Milky Way and discuss possible
interpretations. We consider the most plausible explanation to be a dense
cloud in the foreground of both the X-ray and infrared emitting regions
which obscures the entire X-ray source but only a fraction of the much
larger infrared source.
\end{abstract}
\keywords{galaxies: active --- galaxies: individual (NGC~3281) ---
galaxies: nuclei --- galaxies: Seyfert --- infrared: galaxies ---
X-rays: galaxies}

\section{Introduction}

In the current paradigm for the unification of Seyfert galaxies (see,
e.g.\ Antonucci 1993), Seyfert~1 and Seyfert~2 galaxies are intrinsically
the same object, the observed differences being due to significant
obscuration along the line of sight to the nucleus in the latter class.
This obscuration blocks the broad line region (BLR) and nuclear continuum
source from view, leaving only the narrow lines clearly visible to
indicate the presence of nuclear activity. Evidence to support this
picture has come from many different areas, including spectropolarimetry
(e.g.\ Antonucci \& Miller 1985; Miller \& Goodrich 1990), X-ray
spectroscopy (Awaki et al.\ 1991; Turner et al.\ 1997a, 1997b, 1998), and
near-infrared spectroscopy (Blanco, Ward \& Wright 1990; Goodrich,
Veilleux \& Hill 1994), and it is now almost universally accepted.

The material responsible for obscuring the Seyfert nucleus lies
preferentially in the equatorial plane of the AGN (whose axis is defined
by the radio jets), and for this reason it is commonly referred to as the
``torus''. Goodrich et al.\ (1994) have commented that measuring the line
of sight opacity in a large sample of Seyfert galaxies could allow
information to be gleaned about the torus' geometry. Such measurements can
be made at both near-infrared and X-ray wavelengths; the former
observations determine the amount of dust along the line of sight, while
the latter measure the photoelectric absorption column which can be
converted into an equivalent hydrogen column density if the elemental
abundances are known. Although it is believed that the X-rays are produced
close to the black hole, and the near-infrared radiation comes from hot
dust further out, the lines of sight to the two continuum sources should
be similar since the BLR clouds have a low covering factor ($\lesssim
10$\%; e.g.\ Oke \& Korycansky 1982; Shields, Ferland \& Peterson
1995). Goodrich et al.\ (1994) state that it should therefore be possible
to combine near-infrared and X-ray column measurements in a statistical
manner if the gas-to-dust ratio (or, more correctly, the ratio of
effective hydrogen column density to visual extinction, $\NH/A_V$) along
the line of sight is shown to have the Galactic value of $\NH/A_V = 1.9
\times 10^{25}\m$\,mag$^{-1}$ (Bohlin et al.\ 1978, assuming $R \equiv A_V
/ E(B-V) = 3.1$).

Preliminary evidence favoring a universal value of the $\NH/A_V$ ratio has
come from detailed studies of two active galaxies, IC~5063 (Simpson, Ward
\& Kotilainen 1994) and Cygnus~A (Simpson 1994a; Ward 1996). In both cases
the derived values of $\NH$ and $A_V$ were found to have a similar ratio
to Galactic lines of sight. However, Alonso-Herrero, Ward \& Kotilainen
(1997) have analyzed a sample of Seyfert~2 galaxies and found that the
ratio can often be much larger. Unfortunately, the infrared data available
to Alonso-Herrero et al.\ were of lower quality than those used in the
studies by Simpson and colleagues, and their results are therefore less
certain. High-quality data on a large sample of galaxies should produce a
more conclusive result.

In this paper, we present a thorough near-infrared and X-ray study of the
nucleus of the Seyfert~2 galaxy NGC~3281, and find that the $\NH/A_V$
ratio along our line of sight which is more than an order of magnitude
greater than the Galactic value. We investigate possible interpretations
of this result. Throughout this paper we adopt a recession velocity with
respect to the Galactic Standard of Rest for NGC~3281 of
3224\,km\,s$^{-1}$ (de~Vaucouleurs et al.\ 1991). With a Hubble constant
$H_0 = 50$\,km\,s$^{-1}$\,Mpc$^{-1}$, our assumed distance is 64.5\,Mpc.

\section{The infrared images}

\subsection{Observations}

NGC~3281 was observed using the IRAC-1 infrared array camera on the
European Southern Observatory 2.2\,m telescope at La Silla, Chile, on the
night of UT 1994 Dec 17.  Images were taken in the $J$, $H$, and $K$
filters, with on-source integration times of 180\,s at $J$ and $H$, and
240\,s at $K$. An equal amount of time spent on regions of sky 2\arcmin\
away, and these images were used to flatfield and sky-subtract the images
of the galaxy. Since the conditions were non-photometric during the
observations, we have flux calibrated our images from the aperture
photometry of Glass \& Moorwood (1985). We use their 12\arcsec\ aperture
fluxes, since these will be less affected by possible variability of the
Seyfert nucleus. However, we note that our data have the same ratio of
12\arcsec\ to 6\arcsec\ aperture fluxes as those of Glass \& Moorwood, so
there appears to be no significant variability between the two epochs.

Further infrared images of NGC~3281 were taken with IRCAM3 on the 3.8\,m
United Kingdom Infrared Telescope (UKIRT) at the summit of Mauna Kea,
Hawaii, on UT 1996 Jan 5, a few days prior to our {\it ASCA\/} observation
(\S3.2). Two 60\,s images were taken in the $K$ filter, one centered on
NGC~3281, and one on a nearby blank region of sky. Images were also taken
in the $L'$ and $M$ filters, totaling 300\,s on source and 300\,s on
neighboring patches of sky, to enable flatfielding. Flux calibration was
performed using observations of UKIRT photometric standards over the
course of the night. Simulated aperture photometry of our $K$ image agrees
to within 5\% with that of Glass \& Moorwood (1985).

\subsection{Separation of galaxy and nucleus}

We have used two different procedures to measure the fluxes of a possible
unresolved nuclear source in the infrared images since the relative
contributions of nucleus and host galaxy vary greatly with wavelength. To
determine the nuclear flux at $H$ and $K$ in the ESO images, we subtracted
the ESO $J$ image from the images in these filters, after scaling it so
that the counts matched in an annulus $3\arcsec < r < 5\arcsec$, and then
performed photometry in a 3\arcsec\ aperture on the residual unresolved
source. We opt for this method rather than the profile fitting advocated
by Simpson (1994b) because of the small field of view of IRAC-1 and the
uncertain sky level. The nuclear excess is clearly visible in radial
surface brightness profiles, and the correction needed account for the
presence of color gradients which caused Simpson to disfavor this method
is small for a galaxy as extended as NGC~3281. At $L'$ and $M$, the host
galaxy is not detected with any significance, so we have performed
photometry within a 3\arcsec\ aperture, and corrected our measurements for
contamination from the underlying host galaxy using the spectrum of
$\beta$~Peg (Strecker, Erickson \& Witteborn 1979), an M2~II--III star
whose infrared colors match those of the normal ellipticals in the sample
of Frogel et al.\ (1978). We estimate that the host galaxy contributes
only 10\% and 2\% of the flux within the 3\arcsec\ aperture at $L'$ and
$M$ respectively, and our nuclear magnitudes are therefore insensitive to
the assumed galaxy colors. We have also used the aperture photometry of
Glass \& Moorwood (1985) to determine the strength of the nuclear source
at $L$, performing a similar analysis to that used on the ESO data. We
present the results of the photometry in Table~\ref{tab:irdata}.

\subsection{Extinction to the nucleus}

\begin{figure}
\plotone{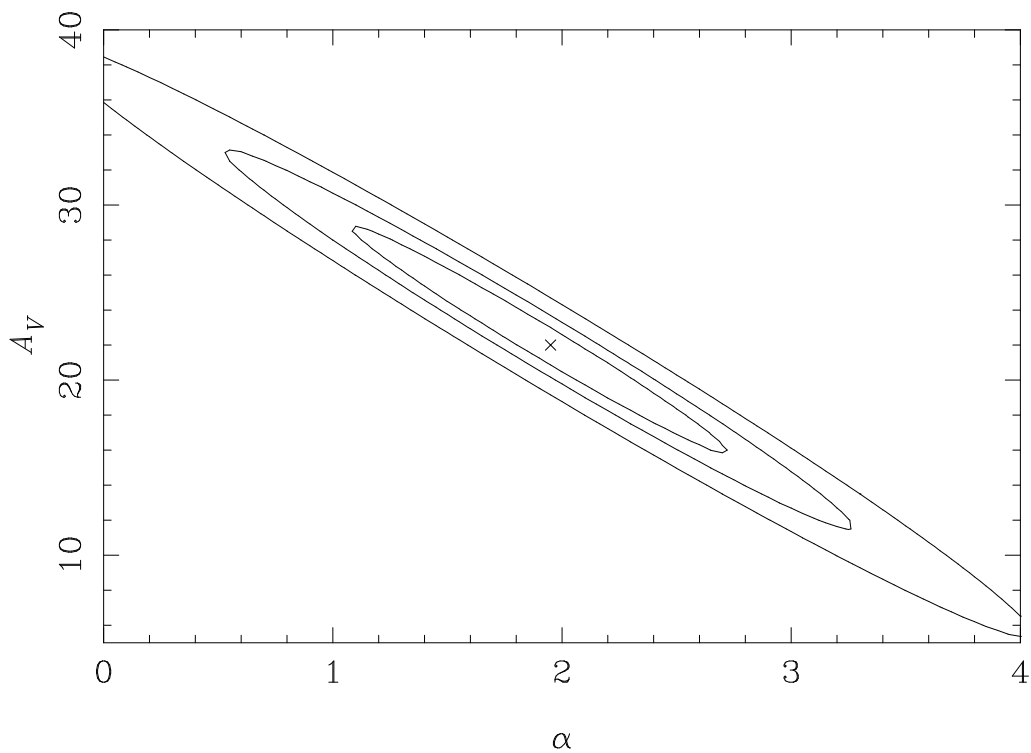}
\caption[avalpha]{Confidence contours in the spectral index--extinction
plane for fitting a reddened power law to the nuclear fluxes. The cross
marks the minimum of $\chi^2$, and the contours are at 68\%, 90\%, and
99\% confidence intervals. Spectral index is defined in the sense $S_\nu
\propto \nu^{\alpha}$.}
\label{fig:avalpha}
\end{figure}

\begin{figure}
\plotone{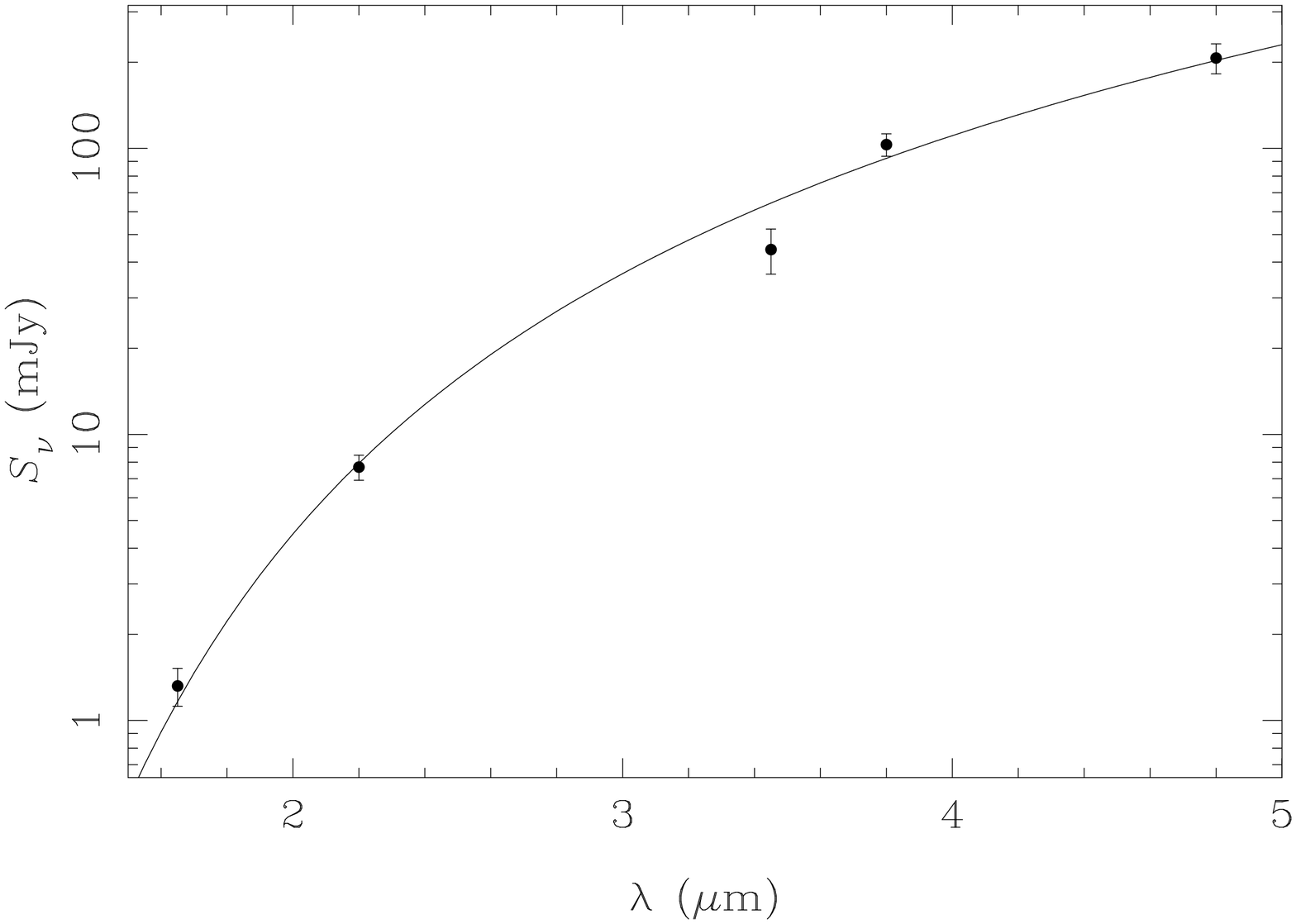}
\caption[ircols]{The measured fluxes of the nuclear source in NGC~3281
(points with error bars) and the best fit reddened power law (solid
line), as determined by the analysis of Figure~\ref{fig:avalpha}.}
\label{fig:ircols}
\end{figure}

Studies of large samples of quasars (e.g.\ Neugebauer et al.\ 1987) have
shown that the near-infrared continuum is a power law with spectral index
$\alpha = 1.4 \pm 0.3$ ($S_\nu \propto \nu^{-\alpha}$), believed to be due
to thermal emission from hot dust at a range of temperatures. Seyfert
galaxies are believed to be merely scaled-down versions of quasars and
their near-infrared continua should be similar. Fadda et al.\ (1998) find
slightly steeper values for Seyfert~1s ($\alpha \approx 1.7$--2.0),
although this is quite critical on the separation of host galaxy and
nucleus at the shorter wavelengths. We assume that the intrinsic
near-infrared continuum of NGC~3281 is also well described by a power law
and perform a grid search in the spectral index--reddening plane to
determine what values of $\alpha$ and the nuclear extinction produces the
best fit to the observed nuclear magnitudes. We use the interstellar
extinction law of Rieke \& Lebofsky (1985), and find that $A_V = 22 \pm
11$ and $\alpha = 1.95^{+1.30}_{-1.40}$ (90\% confidence intervals). As
Figure~\ref{fig:avalpha} shows, the two parameters are strongly
anti-correlated, since increased extinction has the effect of steepening
the spectrum. The best-fitting reddened power law model is shown in
Figure~\ref{fig:ircols}.

An additional consistency check can be made using the observed strength of
the [\ion{O}{3}]~$\lambda$5007 emission line, which is an isotropic
indicator of the strength of the Seyfert nucleus (see Mulchaey et al.\
1994). Since Seyfert galaxies are believed to be merely scaled-down
versions of radio-quiet quasars, we use the equivalent width of 24\,\AA\
(with scatter of a factor 2) for this line in the bright quasar sample
(Miller et al.\ 1992) to estimate the unobscured optical continuum level
for the isolated Seyfert nucleus (since Miller et al.\ observed bright
quasars, the contamination by starlight from the host galaxy will be small
and hence their measurement of the equivalent width is an indication of
the line strength relative to the non-stellar continuum alone). The
integrated flux of $f_{\rm[O\,III]} = 1.0 \times 10^{-15}$\,W\,m$^{-2}$
measured by Storchi-Bergmann, Wilson \& Baldwin (1992b) then implies a
continuum flux at 5007\,\AA\ of $4.2 \times
10^{-17}$\,W\,m$^{-2}$\,\AA$^{-1}$, uncertain by a factor of 2. By
adopting the mean optical--near-infrared quasar spectrum from Neugebauer
et al.\ (1987), we determine an unobscured nuclear magnitude of $K = 9.33
\pm 1.08$, implying an extinction $A_K = 3.00 \pm 1.08$, or $A_V = 27 \pm
10$. The good agreement between this number and our near-infrared color
analysis adds further weight to our interpretation of the nuclear source
in NGC~3281 as a reddened Seyfert nucleus.

The above analysis has implicitly assumed that the emission arises in a
single region of negligibly small optical depth, seen through a foreground
screen. However, since the longer wavelength emission is emitted by cooler
dust located further from the nucleus, it will provide additional
obscuration along the line of sight to the hotter dust closer to the
nucleus. One might therefore expect the foreground screen model to fit the
data poorly, either underestimating the longer wavelength fluxes or
overestimating the shorter wavelength emission. We briefly show here that
the optical depth of the near-infrared emitting region is too small to
have a significant effect.

Following Barvainis (1987), if the dust has a temperature $T_{\rm i}$ at
the inner edge of the torus, radius $r_{\rm i}$, then the optical depth at
ultraviolet wavelengths from the nucleus to some radius $r > r_{\rm i}$
where the dust temperature is $T$ is given by
\[
\tau_{\rm UV} = 5.6 \ln (T_{\rm i} / T) + 2 \ln (r_{\rm i} / r) .
\]
The second term on the right hand side is always negative (since $r >
r_{\rm i})$, so $\tau_{\rm UV} < 5.6 \ln (T_{\rm i} / T)$. The longest
wavelength we have studied is 4.8\,\micron, which is the peak wavelength
for dust at a temperature of 600\,K. We take $T_{\rm i}$ to be 1500\,K
(Barvainis 1987), hence $\tau_{\rm UV} \approx 5$ at the
4.8\,\micron-emitting region. Making the same assumptions about the
heating continuum as Storchi-Bergmann et al.\ (1992a,b), this optical
depth can be achieved with as little as 5\,mag of visual extinction, which
is similar to our uncertainty and much lower than the column we derive. It
is therefore not surprising that a simple foreground screen model provides
an acceptable fit to the observations.

It can also be shown straightforwardly that a collection of sources
emitting the same spectrum, but viewed through different amounts of
extinction, cannot combine to mimic a foreground screen. While it is true
that several sources emitting {\em different\/} spectra, and seen through
different extinctions, could achieve this feat, it is not only contrived
but cannot explain the low dispersion observed in the near-infrared
spectral indices of quasars and Seyfert~1 galaxies. In these objects, the
near-IR emission is seen effectively unobscured and, if this emission is
produced by a number of sources with different spectral shapes, it is
unclear how these could always combine to produce the same overall
spectrum. Rather, the emission must come from a single, coherent region.
Our interpretation of the nuclear source in NGC~3281 as a Seyfert~1-like
nucleus seen through $A_V \approx 22$\,mag of foreground extinction is
therefore the most plausible explanation.

\section{The X-ray data}

\subsection{Previous X-ray observations}

NGC~3281 appears in the {\it Ariel V\/} catalogue (McHardy et al.\ 1981)
as the proposed identification for the source 3A~1030$-$346. Since the
error box is approximately 1.8\,deg$^2$, this identification cannot be
considered secure, although the 2--10\,keV flux of $(2.9 \pm 0.6) \times
10^{-14}$\,W\,m$^{-2}$ implies a ratio of [\ion{O}{3}] to hard X-ray flux
typical of Seyfert galaxies (Mulchaey et al.\ 1994).  However, a source
with a flux of $(3.6 \pm 0.4) \times 10^{-14}$\,W\,m$^{-2}$ was detected
by {\it HEAO\/} A-1 (1H~1027$-$351; Wood et al.\ 1984), and although its
error box overlaps with that of 3A~1030$-$346, it does {\em not\/} include
NGC~3281. Since {\it HEAO\/} A-1 had greater spectral coverage and higher
sensitivity that {\it Ariel V\/}, it should have detected all the sources
in the {\it Ariel V\/} catalog. Given the consistent flux measurements and
positions of 3A~1030$-$346 and 1H~1027$-$351, we propose that these
sources are the same, as yet unidentified (to the best of our knowledge),
X-ray source located within the region of overlap between their two error
boxes. Neither of them should be identified with NGC~3281.

Consistent with this interpretation is a pointed observation made with the
IPC on board the {\it Einstein\/} satellite (Fabbiano, Kim \& Trinchieri
1992). The upper limit to the count rate of 0.017\,counts\,s$^{-1}$ is an
order of magnitude lower than that expected if NGC~3281 has a ratio of
intrinsic 2--10\,keV to [\ion{O}{3}] luminosities typical of Seyferts, and
is seen through a column density $\NH = 5.3 \times 10^{26}\m$
(corresponding to $A_V = 28$\,mag and the $\NH/A_V$ ratio of Bohlin et
al.\ 1978). It is possible that NGC~3281 is a relatively feeble emitter of
X-rays, but since the bandpass of the {\it Einstein\/} IPC is fairly soft
(0.16--3.5\,keV), the observed flux is highly dependent on the amount of
photoelectric absorption, and a large absorbing column might be the
culprit.

\subsection{ASCA Observations and reduction}

The {\it ASCA\/} X-ray satellite was used to observe NGC~3281 on UT 1996
Jan 8, to determine the intrinsic X-ray luminosity and photoelectric
absorbing column. The SIS detectors operated in 1-CCD faint mode, and the
GIS detectors in PH mode. Standard screening criteria were used to exclude
periods of bad data. The raw data were screened to exclude events detected
while the telescope was pointing within 10\arcdeg\ of the Earth, and those
detected within 60\,s of a passage through the South Atlantic Anomaly or
within 60\,s of a day/night transition. Data taken with the GIS detectors
when the cut-off rigidity (COR) was below 7\,GeV/$c$, or with the SIS when
the COR was below 6\,GeV/$c$, were also excluded. Finally, events detected
by the SIS when the telescope was pointing within 20\arcdeg\ of the
sun-illuminated Earth were also excluded. After performing this screening,
light curves were produced which revealed a few short periods of
observation, sandwiched between longer periods of rejected data, when the
background was unstable and unusually high. These short periods of
unstable background, totaling approximately 500\,s, were also excluded
from the analysis. The total exposure times in the four detectors after
screening are listed in Table~\ref{tab:xobs}.

We converted our SIS data to bright mode to allow us to use the archived
high signal-to-noise ratio blank sky observations for background
subtraction, and used the same extraction regions for both the source and
sky. We used the minimum recommended extraction regions (3\arcmin\ for the
SIS detectors, 4\arcmin\ for the GIS detectors) where possible, since the
source was faint and unresolved. For the SIS1 detector, however, this
aperture extended beyond the edge of the CCD and we were therefore obliged
to use a slightly smaller aperture (2\farcm8).

As is the norm for the modeling of X-ray spectra, the observed data were
compared to a model spectrum which had been convolved with the detector
response matrix, using the $\chi^2$ statistic to quantify the quality of
fit. We used the SIS energy response matrices of 1994 Nov 9, and the GIS
response matrices of 1995 Jun 3. In addition, to avoid the difficulties of
error determination when dealing with small number statistics, the
observed data were regrouped so that each bin contained at least 20
events, and channels containing no events were flagged as bad and not used
in determining the best fit. In addition, the GIS channels with energies
below 0.9\,keV and the SIS channels with energies above 8\,keV were also
excluded due to their poor sensitivities and tendency to be affected by
non-random errors.

\subsection{Analysis}

The X-ray continuum of a typical active galactic nucleus (see Mushotzky,
Done \& Pounds 1993) can be modeled as a power law with photon index
$\Gamma \approx 1.9$ plus a reflection hump. In low-quality data such as
ours, the sum of these components resembles a flatter power law with
$\Gamma \approx 1.7$. The spectrum is modified by photoelectric absorption
(both from the Milky Way and the host galaxy) and also features a
K$\alpha$ emission line of iron with a rest energy of 6.4\,keV and an
intrinsic rest frame equivalent width of 100--200\,eV (Mushotzky et al.\
1993; Nandra et al.\ 1997). In addition, there may be a significant
fraction of the continuum which is spatially extended or scattered into
our line of sight, and thus avoids the heavy absorption from the torus
that surrounds the nucleus. In such cases, the spectrum will have two
absorption cutoffs: one at a high energy associated with the torus and one
at a low energy associated with the much smaller gas column through the
host galaxy and Milky Way. NGC~4945 (Done, Madejski \& Smith 1996) is a
good example of such an object, and NGC~3281 possesses a very similar
X-ray spectrum, which we therefore modeled in a similar manner.

\begin{figure}
\plottwo{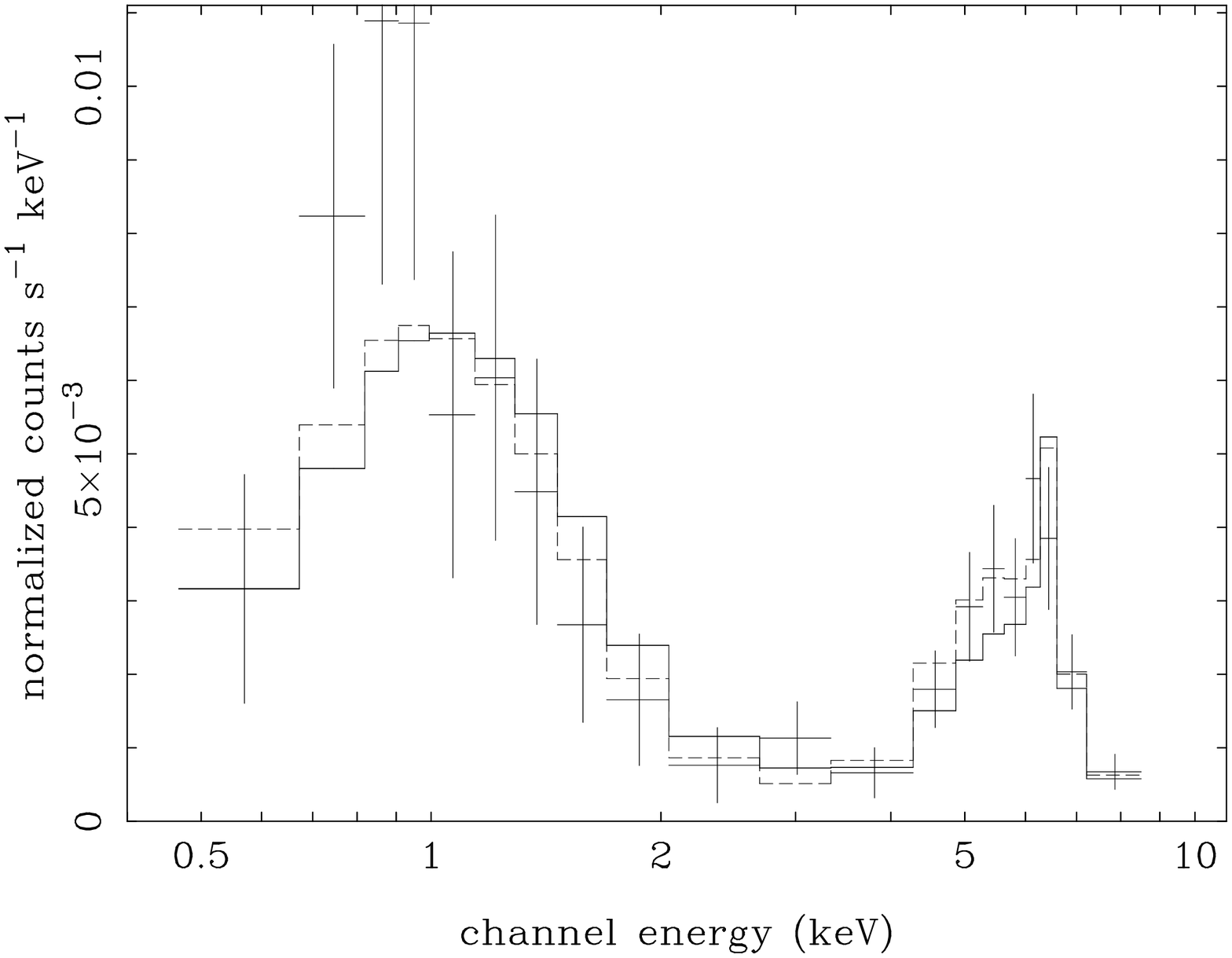}{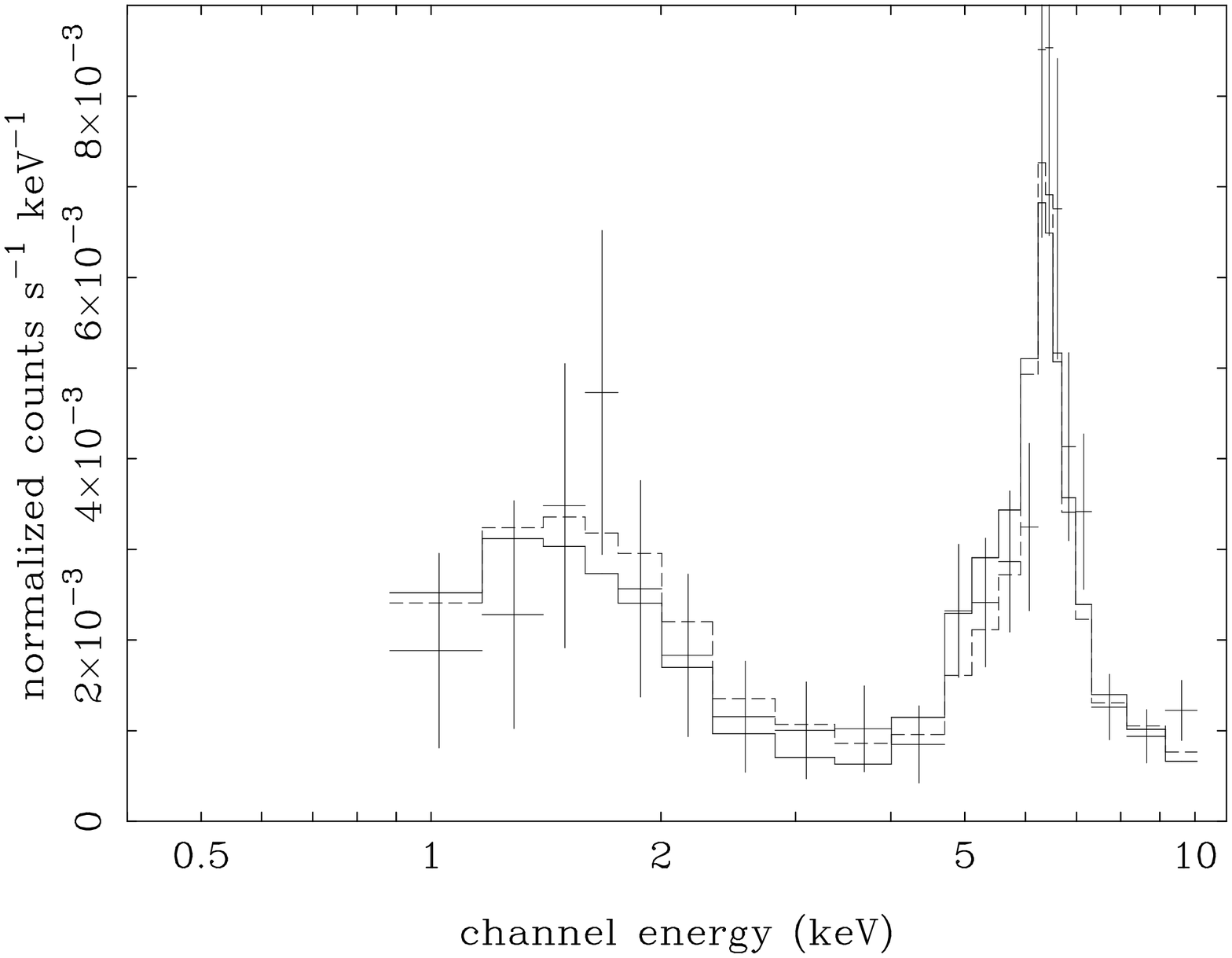}
\caption[xspec]{Data from the SIS0 (left) and GIS3 (right) detectors,
together with the best fit model obtained from fitting (solid lines)
all 4 detectors simultaneously, and (dashed lines) each pair of
detectors only (i.e.\ SIS0+SIS1 or GIS2+GIS3).}
\label{fig:xspecall}
\end{figure}

We performed our analysis with the XSPEC package (Shafer et al.\ 1994),
which uses the Morrison \& McCammon (1983) cross-sections for
photoelectric absorption. Since the power law observed at low energies is
merely an extended or reflected component of the second power law, we
constrained them to have the same photon index. We also fixed the column
density obscuring the scattered/extended continuum at the Galactic value
of $\NH = 6.4 \times 10^{24}\m$ (Heiles \& Cleary 1979). This model gave
an acceptable fit ($\chi^2_\nu = 1.06$ with 73 degrees of freedom),
although the parameters are not particularly well constrained due to the
low count rate. The best fit values for the photon index, equivalent
hydrogen column density along the line of sight to the transmitted
component, and equivalent width of the \FeKa\ line, together with their
associated 90\% confidence intervals are presented in
Table~\ref{tab:xfits}. We show the grouped data and best-fit models for
the SIS0 and GIS3 detectors in Figure~\ref{fig:xspecall}. The \FeKa\ line
is unresolved in all the fits, although our relatively poor
signal-to-noise ratio does not exclude the possibility of a fairly
significant broad component, such as that seen by Turner et al.\ (1998).

\begin{figure}
\plottwo{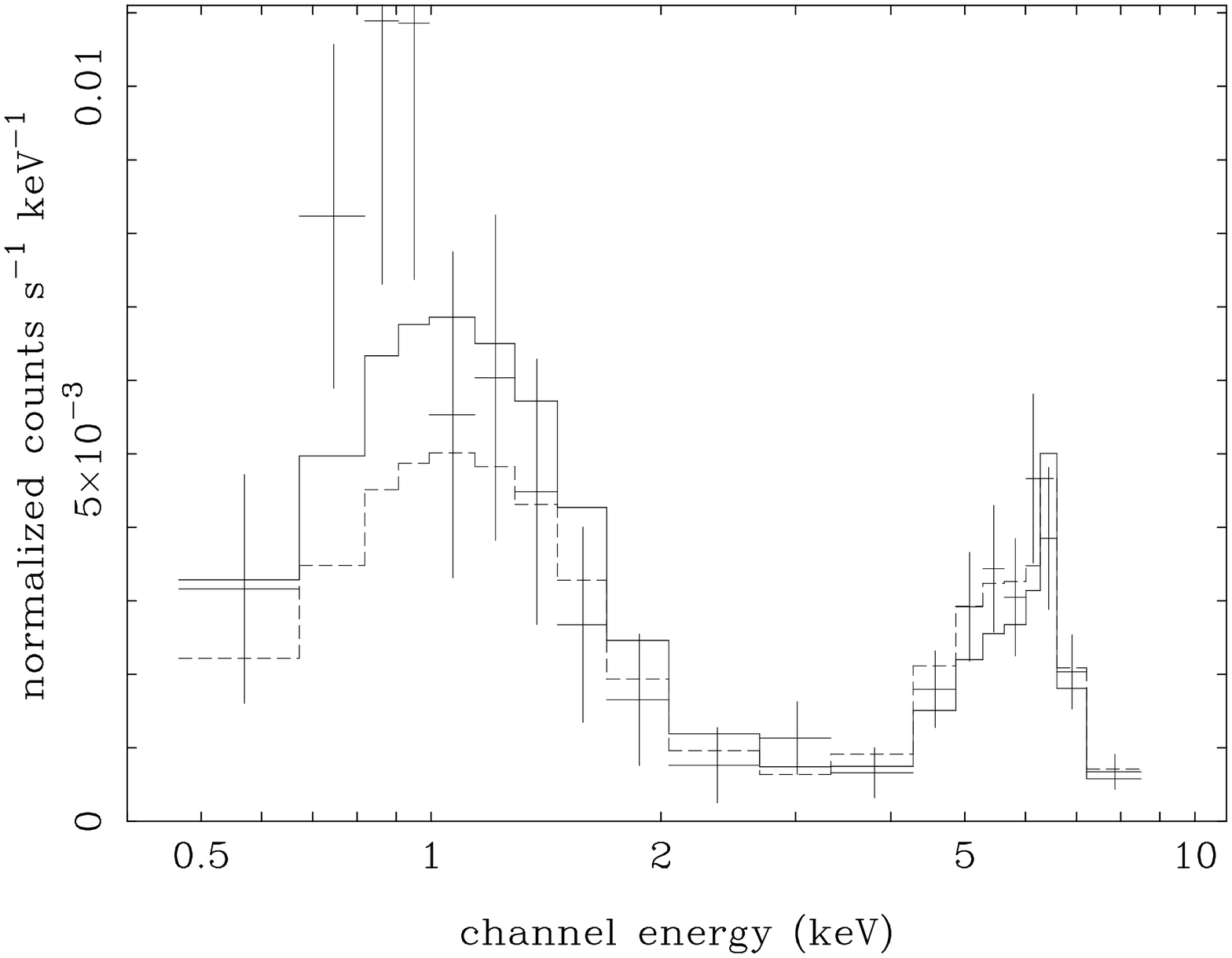}{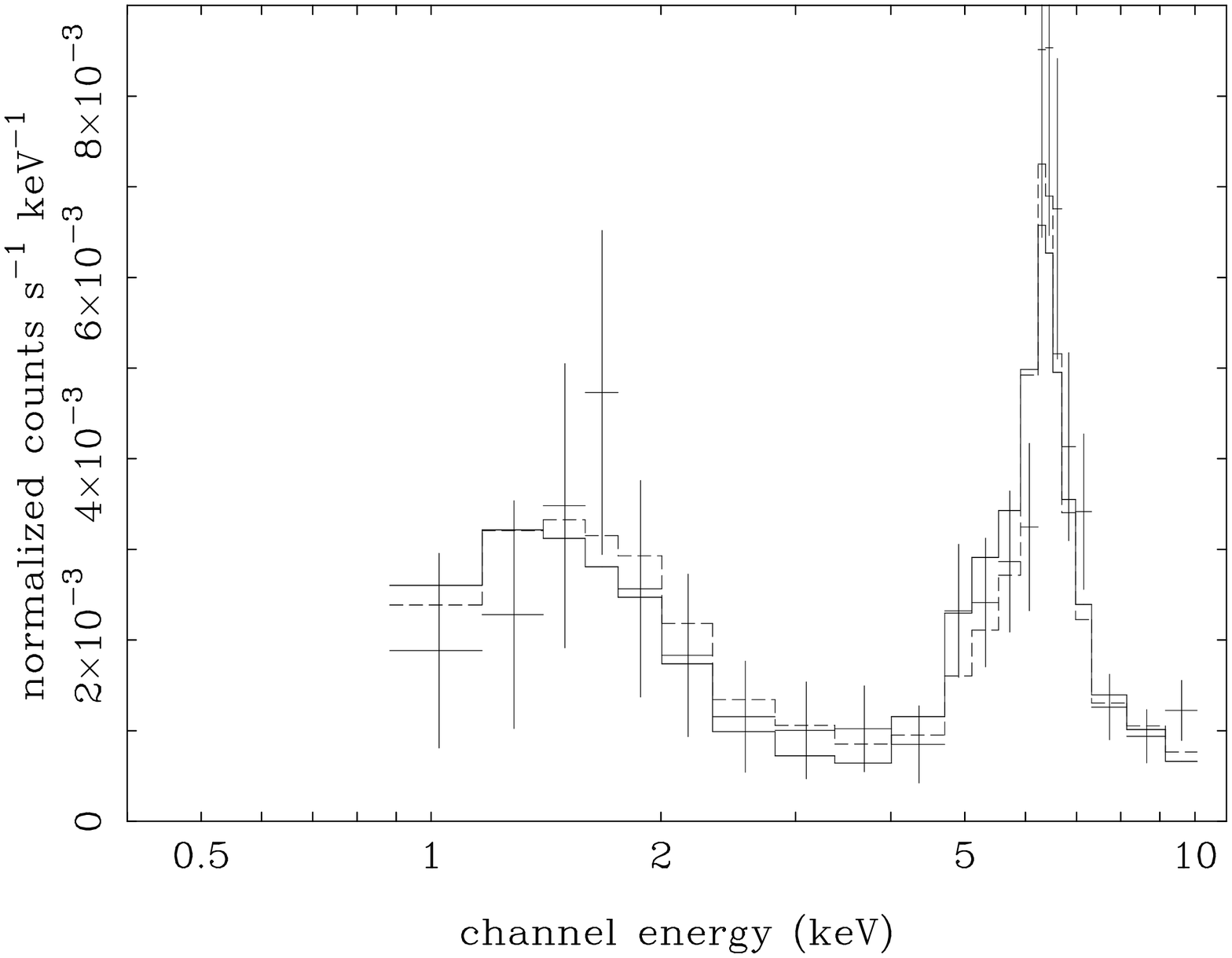}
\caption[xspec]{Data from the SIS0 (left) and GIS3 (right) detectors,
together with the best fit model for the data at energies above 1\,keV
obtained from fitting (solid lines) all 4 detectors simultaneously,
and (dashed lines) each pair of detectors only (i.e.\ SIS0+SIS1 or
GIS2+GIS3).}
\label{fig:xspec1k}
\end{figure}

There is an excess of counts above our best fitting models in both SIS
detectors in the range 0.7--1\,keV. A similar excess has also been seen in
the Seyferts Mrk~3 and NGC~1365 (Turner, Urry \& Mushotzky 1993) and the
LINERs NGC~2639 (Reichert, Mushotzky \& Filippenko 1994) and possibly
NGC~4639 (Koratkar et al.\ 1995). Turner et al.\ (1993) speculatively
attributed this excess to an emission-line complex of Fe~L and ionized
oxygen. Such an explanation might also apply to NGC~3281, although the
sensitivities of the GIS detectors at these energies are insufficient to
determine whether this excess is present in all detectors or only in
SIS0+SIS1, and might therefore be the result of poor background
subtraction in these detectors. Whatever the cause, it is likely that this
excess is artificially increasing the value of the photon index derived
for the SIS spectra, so we have performed additional fitting using only
the data in channels with energies above 1\,keV. These results are also
presented in Table~\ref{tab:xfits} and Figure~\ref{fig:xspec1k}.

\begin{figure}
\plottwo{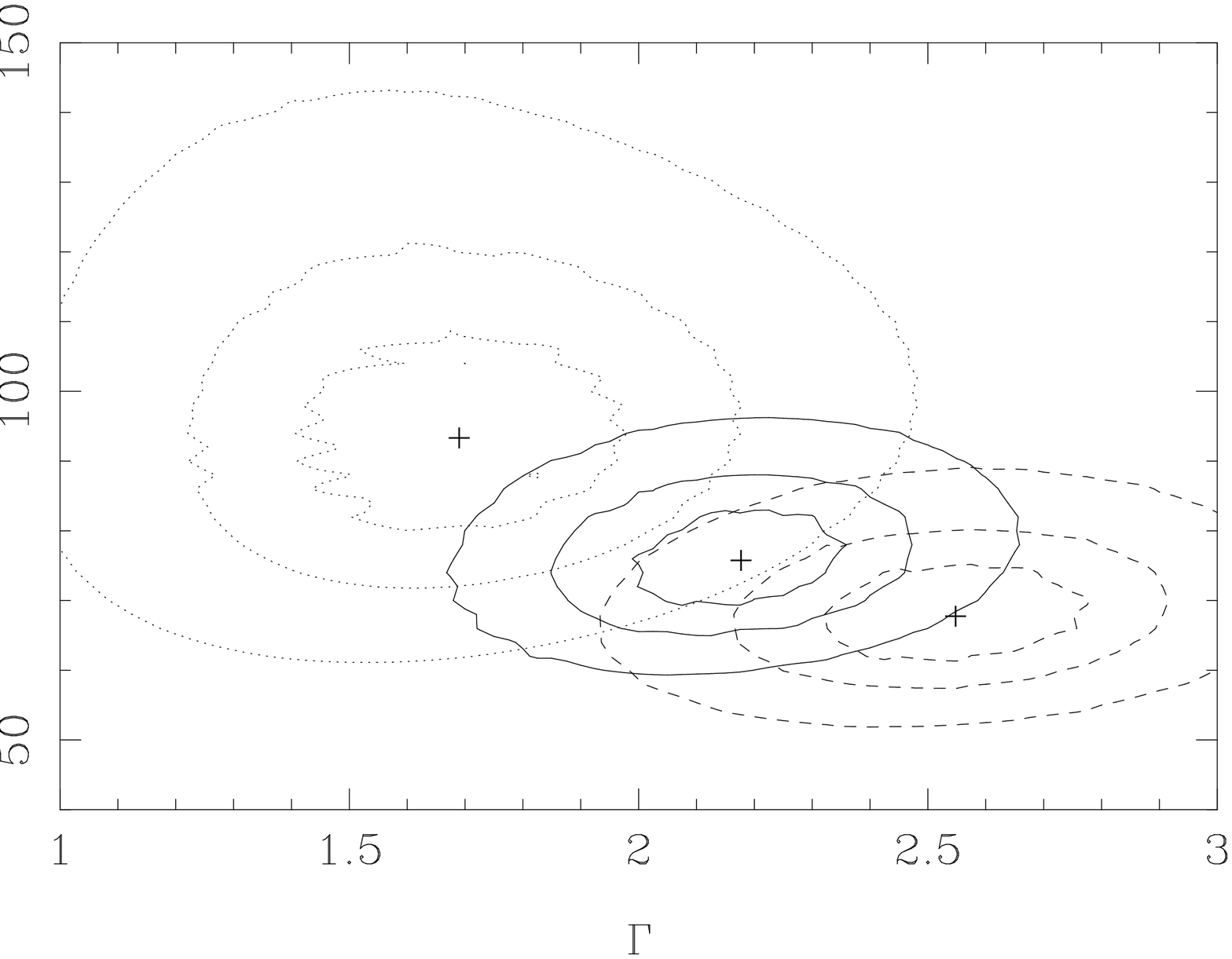}{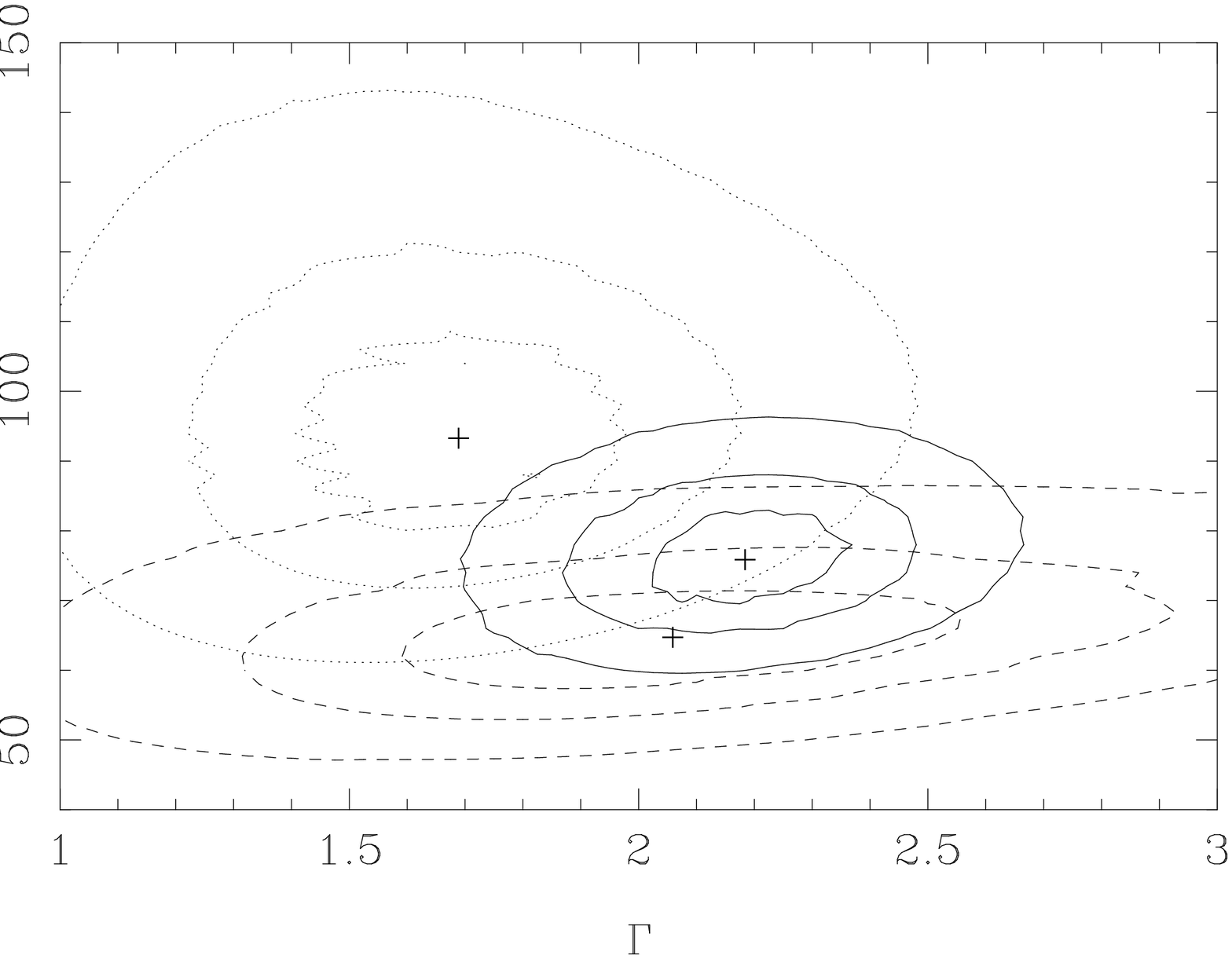}
\caption[xcont]{Photon index--hydrogen column density confidence
contour plots for the entire energy range (left) and energies above
1\,keV only (right). The crosses mark the locations of the best fits
(see Table~\ref{tab:xfits}), and the contours are at 68\%, 90\% and
99\% confidence. Solid lines are confidence contours for the
simultaneous fits to all four detectors, dashed lines for fits to the
SIS0+SIS1 detectors, and dotted lines for the GIS2+GIS3 detectors.}
\label{fig:xcont}
\end{figure}

Figure~\ref{fig:xcont} shows confidence level contour plots for the photon
index and gas column density. It is clear from these figures that the
overall consistency of the fits between the different detectors is
increased when the data below 1\,keV is excluded. We therefore choose to
use the results from fits to the $E > 1$\,keV data only, after noting that
the parameter we are most interested in (the depth of the intervening gas
column) is constrained by the position of the cutoff at $\sim 5$\,keV and
is thus largely unaffected by the inclusion or exclusion of the low energy
data.

We investigated alternative models to see whether they were able to
provide acceptable fits to our data. We added a Compton reflection hump,
which provides additional flux at $E \gtrsim 5$\,keV and might therefore
help to explain the change in flux which we have attributed to
photoelectric absorption. This made no significant change to any of the
fitting parameters, and in fact led to a worse quality of fit indicator by
virtue of the many additional free parameters. We also added a thermal
plasma, based on the calculations of Raymond \& Smith (1977), as it could
be responsible for the putative emission lines at soft energies. We find,
however, that such a plasma does not provide significant flux in the
1--3\,keV range, and contributes negligibly at higher energies. It can
therefore have little or no effect on the hydrogen column density we
derive, which Table~\ref{tab:xfits} shows remains fairly constant over a
large range of photon indices.

The observed 2--10\,keV flux of the best fitting model is $2.7 \times
10^{-15}$\,W\,m$^{-2}$, more than an order of magnitude fainter than the
source(s) detected by {\it Ariel~V\/} and {\it HEAO\/} A-1. This supports
the assertion that the detections made by these satellites are of a
separate and presumably unrelated source. The intrinsic 2--10\,keV flux
(i.e.\ in the absence of the large absorbing column) is $2.3 \times
10^{-14}$\,W\,m$^{-2}$, so the ratio of [\ion{O}{3}] to hard X-ray flux is
within the $1\sigma$ scatter for Seyfert galaxies (Mulchaey et al.\ 1994),
and the absorption-corrected luminosity $L_{\rm2-10\,keV} = 1.1 \times
10^{36}$\,W is typical of Seyfert galaxies. The few other Seyfert~2s known
with similarly large absorption columns ($\NH \gtrsim 10^{28}\m$; e.g.\
NGC~1068, NGC~4945) show large \FeKa\ equivalent widths, apparently
because the continuum photons are preferentially absorbed over the line
photons. Although the uncertainties on the equivalent width of the \FeKa\
line are large, our derived value of $\sim 500$\,eV is larger than typical
values, and so NGC~3281 falls naturally into this picture. Our model
therefore not only provides an acceptable fit to the data, but also
suggests that the intrinsic X-ray spectrum of NGC~3281 is that of a
typical Seyfert galaxy.

\section{Discussion}

In \S2 we showed that the near-infrared properties of the unresolved
nuclear source in NGC~3281 are consistent with a normal Seyfert~1 spectrum
seen through $A_V = 22 \pm 7$\,mag ($1\sigma$ error) of extinction. In \S3
we showed that NGC~3281's X-ray properties are consistent with a normal
Seyfert~1 spectrum seen through an equivalent hydrogen column of $\NH =
(7.1 \pm 1.2) \times 10^{27}\m$ ($1\sigma$ error). Taken together, the
ratio of gas to dust columns, $\NH/A_V = (3.2 \pm 1.7) \times
10^{26}\m$\,mag$^{-1}$ (90\% confidence), is more than an order of
magnitude larger than the Galactic ratio derived by Bohlin et al.\ (1978)
of $1.9 \times 10^{25}\m$\,mag$^{-1}$. The implication of this is fairly
obvious, i.e.\ that extinction estimates from near-infrared and X-ray
analyses cannot be combined in the manner hoped for by Goodrich et al.\
(1994).

We doubt that the apparent dearth of dust along our line of sight to the
nucleus is because NGC~3281 has a low dust content on a galactic scale.
There is significant dust obscuration in the central 20\,kpc of the galaxy
(Storchi-Bergmann et al.\ 1992b), the stellar population of the bulge is
typical of an early-type galaxy and the metallicity near the nucleus is
estimated to be about twice solar, indicating a fairly normal course of
stellar evolution. While it is true that our line of sight through the
thick molecular torus in NGC~3281 is very unlike the Galactic lines of
sight along which Bohlin et al.\ (1978) derived their $\NH/A_V$ ratio, the
analyses of IC~5063 (Simpson et al.\ 1994) and Cygnus~A (Simpson 1994a)
did not produce large $\NH/A_V$ ratios. We cannot therefore make an
arbitrary appeal to unusual physical conditions to explain the
discrepancy, and must look for an alternative explanation.

We note that the high inclination of NGC~3281 means that the obscuring
material may be part of a larger-scale structure in the plane of the
galactic disk (McLeod \& Rieke 1995; Simcoe et al.\ 1997). This does not
affect the following discussion in any way, which relates to the
quantities of gas and dust along the line of sight, and not to their
locations.

\subsection{An alternative origin for the hot dust}

We first rule out the possibility that the near-infrared source we observe
has some other origin than a reddened Seyfert~1 nucleus. The observed
1--5\,\micron\ luminosity of the source is $5 \times 10^9\Lsun$, but its
spectral index ($\alpha \approx 5$) is almost certainly too steep to be
intrinsic, and the reddening requires that the true luminosity be higher.
Blue intrinsic spectra, such as those appropriate for the Rayleigh-Jeans
tail of a starburst, require a luminosity in excess of $10^{12}\Lsun$ in
this wavelength region alone, and so the spectrum must be intrinsically
red, or cool, to avoid such unreasonably large numbers. Thermal emission
from dust is the only plausible candidate, requiring a luminosity of about
$2 \times 10^{10}\Lsun$, which must also be the luminosity of the heating
source. A powerful starburst is the only phenomenon, other than an active
galactic nucleus, which can produce this quantity of radiation on a small
scale, but while starbursts are frequently seen in active galaxies, they
tend to exist in kpc-scale rings, and the unresolved nature of the source
we observe requires the burst to be occurring on a scale an order of
magnitude smaller, despite having a comparable luminosity. A more serious
problem with a starburst as the heating source is that the dust must lie
close to the individual stars for it to be heated to temperature of
$\gtrsim 1000$\,K. A single dust shell around a cluster of a few million O
stars will not work. Yet we know that the good fit we obtain to the data
with a simple foreground screen model precludes a geometry where the dust
is clumped around the stars and suffers a large range of extinctions.

\subsection{Supersolar metallicity}

Although we quote equivalent hydrogen columns, our measurement of the
photoelectric absorption column has been determined by the cutoff at $\sim
5$\,keV, where the total photoelectric absorption cross section is
dominated by heavy elements. The optical depth at this energy is therefore
a measure of the metal abundance along the line of sight, and by assuming
solar metallicity we overestimate the true hydrogen column density, $N_H$,
by a factor approximately equal to the metallicity (relative to
solar)\footnote{Most of the opacity is provided by O, Si, S and the Fe
L-shell, so it is the abundances of these elements which most strongly
determine the effective column density.}. If the abundances of these
elements are enhanced by a factor of $\sim 10$ above solar, the {\em
true\/} hydrogen column density would be in line with the near-infrared
extinction.

It is known that some heavy elements (most notably N and O) have high
abundances in the centers of early-type spirals (e.g.\ Storchi-Bergmann,
Wilson \& Baldwin 1996), but they are typically no more than a factor of
3--4 above solar. In addition, other elements need not be overabundant
since the enrichment of nitrogen and oxygen is usually explained by the
CNO cycle. Quadrupling the N and O abundances alone increases the
photoelectric absorption cross-section at 6.4\,keV by less than a factor
of two. While the very high metallicity required need only exist within
the torus, it is unclear how such a confinement might occur, and more
importantly why a similar effect was not observed in either IC~5063 or
Cygnus~A. While we may have overestimated the hydrogen column density by a
factor of about two (spectroscopy of the narrow line gas indicates $Z
\approx 2 Z_\odot$; Storchi-Bergmann et al.\ 1992b), we have not done so
by the order of magnitude necessary to bring it in line with the dust
column.

\subsection{A dense obscuring cloud}

Since both columns we have measured appear to be true representations of
the amount of material along the lines of sight, we explore a geometric
interpretation which exploits the fact that the near-infrared and X-ray
sources are not cospatial. Perhaps the most natural explanation is that a
BLR cloud lies along the line of sight to the X-ray emitting region. This
could contribute significantly to the photoelectric absorbing column, but
would not affect the extinction to the hot dust since the BLR is located
closer to the nucleus and therefore not in the dust's foreground.
Unfortunately, the time for a cloud with velocity $10^4$\,km\,s$^{-1}$ to
traverse the BLR is only about one year (Wanders et al.\ 1995), yet the
extra photoelectric absorption must have been present at the times of the
three previous X-ray observations. The low covering factor of BLR clouds
appears to exclude the possibility of a transient phenomenon like this
occurring at four separate observational epochs. On the other hand, if BLR
clouds have a larger covering factor cover a small range of solid angle
(e.g.\ close to the equatorial plane), and our line of sight passes
through this region, the probability would be larger. For the probability
of a BLR cloud to lie along our line of sight on four separate occasions
to be significant, however, the probability of a cloud to obscure the
X-ray nucleus on any single occasion must be fairly large, and therefore
we should expect analyses such as the one we perform here to frequently
produce large $\NH/A_V$ ratios, again at odds with the results for IC~5063
and Cygnus~A.

\begin{figure}
\plotone{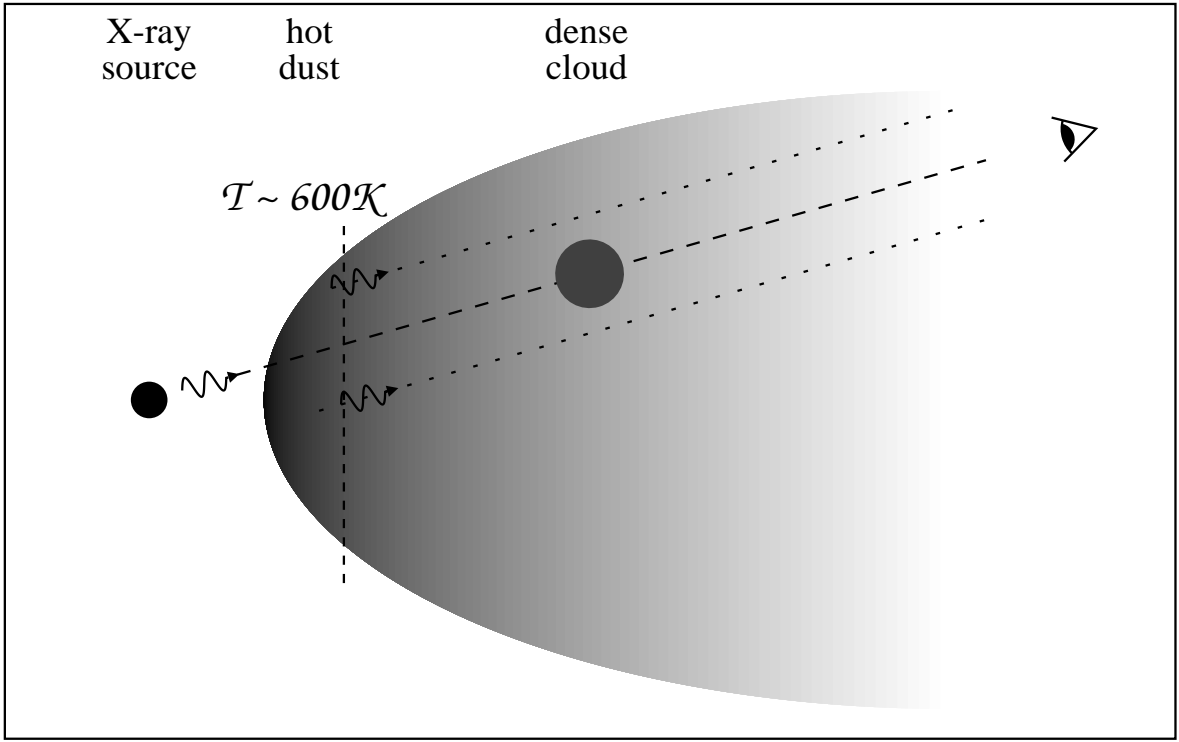}
\caption[cloud]{Schematic diagram showing how a dense cloud located within
the torus can explain the very large $\NH/A_V$ ratio we measure in
NGC~3281. The line of sight to the X-ray source (long dashed line) passes
through a dense cloud with a high column density, while much of the
infrared-emitting hot dust suffers much lower obscuration (short dashed
lines).}
\label{fig:cloud}
\end{figure}

A cloud located in front of both emission regions can provide an
explanation. Since the X-rays come from a much smaller region than do the
near-infrared photons, a cloud could completely cover the X-ray source and
yet still permit an unobscured view of much of the infrared-emitting
region. If such a cloud were sufficiently optically thick to block out all
near-infrared radiation beyond it, even at 5\,\micron, we would only
detect the emission from the unobscured regions, and hence derive a low
extinction. This situation is shown schematically in
Figure~\ref{fig:cloud}.

It is clear that for this picture to work, the cloud must be smaller than
the size of the infrared-emitting region, which is set by the height of
the inner regions of the torus. The inner radius of the torus, $r$, is
determined by the location at which dust cannot exist in thermal
equilibrium with the nuclear radiation field, and so its height, $h$, is
given by
\[
h \approx 2 r \cot \theta_{\rm c} = 0.5 (\pi \sigma)^{-0.5} L^{0.5}
(Q_{\rm a}/Q_{\rm e})^{0.5} T_{\rm sub}^{-2} \cot \theta_{\rm c} ,
\]
where $\theta_{\rm c} = 35\arcdeg$ is the half-opening angle of the
ionization cone (Storchi-Bergmann et al.\ 1992b), $L = 8 \times
10^{36}$\,W is the heating luminosity of the nucleus, $Q_{\rm a}/Q_{\rm e}
= 100$ is the ratio of the dust optical absorption to infrared emission
coefficients and $T_{\rm sub} = 1500$\,K is the sublimation temperature of
dust. The torus is therefore $\sim 0.7$\,pc in height, and for a spherical
cloud to be smaller than this yet still produce most of the observed
photoelectric absorption, its density must be $n_{\rm H} \gtrsim 3 \times
10^{11}\,{\rm m}^{-3}$. Although large, this is only two orders of
magnitude larger than the mean density needed to produce a Compton-thick
torus smaller than 100\,pc, and it is therefore not unreasonable to
believe that such dense condensations could exist. In fact, it might be
more accurate to consider such condensations as inhomogeneities, rather
than {\em bona fide\/} clouds.

Because this cloud lies further from the nucleus than do the BLR clouds,
it will be moving more slowly and the increased absorption will persist
over a much longer timescale,
\[
t \sim 1300 \left( \frac{d_{\rm c}}{0.7\,{\rm pc}} \right) \left(
\frac{v}{300\,{\rm km\,s}^{-1}} \right)^{-1} \,{\rm yr} .
\]
It is therefore entirely plausible that such a cloud could have obscured
the X-ray source during all four observations. Indeed, we would expect the
strong X-ray absorption to persist for many years to come.

\section{Summary}

We have presented a detailed near-infrared and X-ray analysis of the
Seyfert~2 galaxy NGC~3281. We have shown that its near-infrared and X-ray
properties, taken separately, are unremarkable and support the standard
unification model for Seyfert galaxies where Seyfert~2s are simply
Seyfert~1 galaxies seen through significant obscuration. However, the
ratio of the gas and dust column densities we derive is $\NH/A_V = (3.2
\pm 1.7) \times 10^{26}\m$\,mag$^{-1}$ (90\% confidence), more than an
order of magnitude larger than the ratio derived by Bohlin et al.\ (1978)
for lines of sight within the Milky Way. It is clear, therefore, that
obscuring columns derived from near-infrared and X-ray observations cannot
be arbitrarily interchanged using a normal gas-to-dust ratio, as Goodrich
et al.\ (1994) had suggested.

Although we cannot rule out unusual physical conditions in the torus of
NGC~3281 as the cause for the large ratio, such conditions are not common
to all tori, since previous studies of two other active galaxies (Simpson
et al.\ 1994; Simpson 1994a) produced results consistent with the Galactic
ratio. Among the alternatives we have investigated, we have ruled out a
compact starburst and a BLR cloud lying along the line of sight. We cannot
rule out high metallicity confined to the obscuring material, but consider
it unlikely based on the IC~5063 and Cygnus~A results. Our preferred
explanation is an optically thick cloud which obscures the entire X-ray
source but only a fraction of the infrared source.

We are currently undertaking similar studies of many more Seyfert~2
galaxies in an attempt to learn more about the gas and dust obscuration.
We have disfavored certain scenarios by virtue of their being unable to
explain the IC~5063 and Cygnus~A results, which we consider ``normal'',
yet with a sample of only three objects, it is perhaps premature to
discuss what is and is not normal. In our preferred picture, we would
expect most objects to possess a $\NH/A_V$ ratio similar to the Galactic
value, with the incidence of galaxies with large $\NH/A_V$ ratios
indicating the covering factor of the putative dense clouds, but a larger
dataset may support a different picture. We will report on the results of
this study in the near future.

\acknowledgments

This work has been supported by NASA grant number NAG 5-3393 from the {\it
ASCA\/} Guest Observer Program. The United Kingdom Infrared Telescope is
operated by the Joint Astronomy Centre on behalf of the U. K. Particle
Physics and Astronomy Research Council. Count rates for the {\it
Einstein\/} IPC were determined using the program PIMMS, obtained through
the High Energy Astrophysics Science Archive Research Center Online
Service, provided by the NASA-Goddard Space Flight Center. The author is
extremely grateful to Andrew Wilson and Martin Ward for discussions during
the writing of this paper, and to the referee, Mike Eracleous, for
comments which have substantially improved its content. Parts of this work
were performed at the Jet Propulsion Laboratory, California Institute of
Technology, under a contract with the National Aeronautics and Space
Administration.

\clearpage

\begin{table}
\caption[irdata]{Near-infrared magnitudes for NGC~3281.}
\label{tab:irdata}
\begin{center}
\begin{tabular}{lcrrr@{ }c@{}r}
\tableline \tableline
Telescope/Filter & \multicolumn{1}{c}{$\lambda$ (\micron)} & $m(6\arcsec)$
& \multicolumn{1}{c}{$m_{\rm nucleus}$} & \multicolumn{3}{c}{$S_{\rm
nucleus}$ (mJy)} \\
\tableline
ESO/$J$    & 1.25 & 12.36 & \\
ESO/$H$    & 1.65 & 11.46 & $14.72\pm0.17$ & 1.3 & $\pm$ & 0.2 \\
ESO/$K$    & 2.20 & 10.88 & $12.33\pm0.10$ & 7.7 & $\pm$ & 0.8 \\
SAAO/$L$\tablenotemark{a} & 3.45 & 9.11 & $9.54\pm0.18$ & 44.3 & $\pm$ &
8.0 \\
UKIRT/$K$  & 2.20 & 10.82 & \\
UKIRT/$L'$ & 3.80 &  8.35 &  $8.47\pm0.09$ & 103.1 & $\pm$ & 9.3 \\
UKIRT/$M$  & 4.80 &  7.24 &  $7.27\pm0.12$ & 207.1 & $\pm$ & 24.9 \\
\tableline
\end{tabular}
\tablenotetext{a}{From Glass \& Moorwood (1985).}
\end{center}
\end{table}

\begin{table}
\caption[xobs]{ASCA observations of NGC~3281.}
\label{tab:xobs}
\begin{center}
\begin{tabular}{lccc}
\tableline \tableline
Detector & Exposure time (s) & Source count rate (s$^{-1}$) & Background
count rate (s$^{-1}$) \\
\tableline
SIS0 & 14\,933 & $0.0189 \pm 0.0015$ & $0.0123 \pm 0.0004$ \\
SIS1 & 14\,869 & $0.0156 \pm 0.0015$ & $0.0110 \pm 0.0003$ \\
GIS2 & 15\,310 & $0.0166 \pm 0.0015$ & $0.0092 \pm 0.0002$ \\
GIS3 & 15\,342 & $0.0204 \pm 0.0017$ & $0.0108 \pm 0.0002$ \\
\tableline
\end{tabular}
\end{center}
\end{table}

\begin{table}
\caption[xfits]{Fits to the X-ray data. Uncertainties quoted are 90\%
confidence intervals.}
\label{tab:xfits}
\begin{center}
\begin{tabular}{lccc}
\tableline \tableline
Parameter & All & SIS0+SIS1 & GIS2+GIS3 \\
\tableline
\multicolumn{4}{c}{All data} \\
\tableline
$\chi^2 / \nu$          & 63.57 / 68 & 23.68 / 30 & 20.29 / 34 \\
$\Gamma$                & $2.18^{+0.30}_{-0.29}$
& $2.55^{+0.36}_{-0.37}$ & $1.67^{+0.45}_{-0.48}$ \\
$\NH$ ($10^{26}\m$)      & $75.8^{+11.6}_{-10.6}$
& $67.8^{+9.0}_{-10.1}$ & $93.3^{+28.4}_{-22.8}$ \\
EW$_{{\rm Fe~K}\alpha}$ (eV) & $430^{+470}_{-210}$
& $320^{+400}_{-190}$ & $730^{+1660}_{-480}$ \\
\tableline
\multicolumn{4}{c}{$E > 1$\,keV} \\
\tableline
$\chi^2 / \nu$          & 50.66 / 59 & 12.36 / 21 & 20.07 / 33 \\
$\Gamma$                & $1.82^{+0.37}_{-0.26}$ &
$2.06^{+0.87}_{-0.72}$ & $1.69^{+0.55}_{-0.46}$ \\
$\NH$ ($10^{26}\m$)      & $72.5^{+12.5}_{-12.0}$ &
$64.7^{+13.3}_{-11.6}$ & $88.9^{+26.8}_{-21.3}$ \\
EW$_{{\rm Fe~K}\alpha}$ (eV) & $480^{+770}_{-230}$
& $260^{+900}_{-230}$ & $730^{+2100}_{-500}$ \\
\tableline
\end{tabular}
\end{center}
\end{table}

\clearpage

\end{document}